\documentstyle[aps,prbbib,twocolumn,epsf]{revtex}
\begin{document}

\draft
\title{Asymmetric double quantum dots as a subminiature mesoscopic cell}
\author{{\ Qing-feng Sun$^1$, Jian Wang$^2$, Tsung-han Lin$^1$}}
\address{$^1$State Key Laboratory for Mesoscopic Physics and\\
Department of Physics,\\
Peking University, Beijing 100871, China\\
$^2$ Department of Physics, The University of Hong Kong, Pokfulam Road,\\
Hong Kong, China }
\date{}
\maketitle

\begin{abstract}
A subminiature mesoscopic cell, consisting of an asymmetric double quantum
dot capacitively coupled to a nearby mesoscopic circuit, is proposed, which
can transform disordered noise energy to ordered electric energy. Two
schemes, the noise originating from the nearby mesoscopic circuit and from
the electromagnetic wave disturbance in external environment, are
investigated. We found that the proposed cell can manifest as a good
constant current source and the output current may not reach its largest
value even if the circuit is shorted.
\end{abstract}
\par
PACS numbers: 73.40.Gk, 73.50.Lw, 73.50.Td
%\newpage
\par
Mesoscopic physics, a new branch of condensed matter physics, has been
developed and become an active field in the last two decades, not only
because of the fundamental physics, but also because of the great potential
as regards the new generation of electronic and photonic devices.\cite{ref1}
Very recently, by using inelastic transitions in a tunable two-level system,%
\cite{ref2} Aguado and Kouwenhoven proposed a detector of high-frequency
quantum noise.\cite{ref3} This detector consists of a double quantum dot
which is capacitively coupled to a nearby mesoscopic conductor. Then from
the inelastic current through the double quantum dots, one can measure the
noise spectrum in the nearby conductor.

Noise always exists in the electronic systems, whether Johnson-Nyquist noise
in equilibrium states or shot noise in non-equilibrium states, and has been
investigated extensively.\cite{ref4,ref5,ref6} In addition, a system usually
exists in an electromagnetic environment from that the noise can also be
induced in the system from the electromagnetic wave (EMW) disturbance of the
environment.

In this letter, we propose a subminiature mesoscopic cell, consisting of an
asymmetric double quantum dot (DQD) and a noise offering circuit,
capacitively coupled to a nearby mesoscopic circuit via capacitors $C_c$
(Fig.1(b)), by which one can obtain the useful ordered energy from the
disordered noise energy. In the proposed structure, the asymmetric DQD is
described as a system of three tunnel barriers (See Fig.1(a)) with
capacitances $C$. $Z_{a}$ is the impedance of the appliance load. In this
model, the current noise $S_I$ of the noise offering circuit will give rise
to the fluctuation of the electric potential of the two quantum dot,\cite
{ref3} following the intro-dot energy levels $E_L$ and $E_R$ fluctuate too.
This leads that the electron in the level $E_R$ of the right dot maybe
absorb a quanta from the noise offering circuit and jump to the level $E_L$
of the left dot (see Fig.1(a)). Therefore, a inelastic current goes through
the DQD system from the right side to the left side. We emphasize that no
battery exists in this device, but the system can produce ordered current,
in other words, the system play a role of electric cell. Specifically, we
consider that: (1) Each dot of the DQD has only one electronic level, $E_L$
for the left dot and $E_R$ for the right dot, respectively. (2) The DQD is
designed asymmetric to make the two levels $E_L$ and $E_R$ not in line.
Notice that no gate voltage applied in this DQD for controlling the
intra-dot levels, different from the cases in Ref.\onlinecite{ref7,ref8}.
(3) $E_L< \mu_{L}$ ,$\mu_{R}< E_R$, where $\mu_\alpha$ is the chemical
potential, so that the level $E_{L}$ is almost empty and the level $E_R$ is
almost occupied by the electron. (4) $\Delta E \equiv E_L-E_R \gg T_c$ with $%
T_c$ the coupling between two dots.\cite{ref3} Then the coherent tunneling
between the two dots can be neglected,\cite{ref9} while the incoherent
tunneling dominants, in which the photon emission or absorption to or from
the environment may happen. (5) Let $\Gamma_{L}$ and $\Gamma_{R}$ denote the
tunneling rates through the left and right barriers, and $%
\Gamma_{i}(\epsilon )$ the inelastic transition rate through the middle
barrier. Here we consider the situation with $\Gamma_{i}\ll
\Gamma_{L},\Gamma_{R}$\ , i.e., the middle barrier is much larger than left
and right barriers. Then the inelastic current through the DQD system (from
the right side to the left side) is mainly decided by $\Gamma_{i}(\epsilon )$%
.\cite{ref3} From perturbation theory, one can obtain this inelastic current
as:\cite{ref10,ref11} 
\[
I(\Delta E )=\frac{e}{\hbar }\Gamma_{i}(\Delta E)  f_{L}(E_L)\left[
1-f_{R}(E_R)\right] 
\]
\begin{equation}
-\frac{e}{\hbar }\Gamma_{i}(-\Delta E )f_{R}(E_R)\left[ 1-f_{L}(E_L)\right],
\end{equation}
% \begin{equation}
% I(\Delta E )=\Gamma_{i}(\Delta E) f_{L}(1-f_{R})
% -\Gamma_{i}(-\Delta E )f_{R}(1-f_{L})
% \end{equation}
where $f_{L/R}(\epsilon)=\left\{ e^{(\epsilon -\mu_{L/R})/k_B T_{D}}+1
\right\}^{-1}$ is the Fermi distribution function in the left/right lead, $%
T_{D}$ is the temperature of DQD's system. Here $\Gamma_{i}(\epsilon )$ is
related to the coupling strength $T_{c}$ by $\Gamma_{i}(\epsilon
)=T_{c}^{2}P(\epsilon )$\cite{ref3} and the probability $P(\epsilon )$ for
the exchange of energy quanta with the environment is given (at $\epsilon 
\not{= }0$):\cite{ref3} $P(\epsilon )=\frac{2\pi}{R_K}|Z(\omega)|^{2}
S_{I}(\omega)/\epsilon^{2}$, $S_{I}(\omega)$ is the current noise spectrum
at frequency $\omega$ ($\omega=\epsilon/\hbar$) in the mesoscopic device, $%
R_K=h/e^2 \approx 25.8k\Omega$ is the quantum resistance. $Z(\omega)$ is the
trans-impedance connecting the DQD system and the noise offering circuit.
For the circuit of Fig.1(b), $Z(\omega )$ is: 
\begin{equation}
Z(\omega )=\frac{C_{x}/C}{iC_{x}\omega +\tilde{Z}_{a}^{-1}+\tilde{Z}%
_{s}^{-1}\left( 1+\frac{iC_{x}\omega +\tilde{Z}_{a}^{-1}}{iC_{c}\omega }%
\right) },
\end{equation}
where $\tilde{Z}_{a}=Z_{a}+1/iC\omega $, $\tilde{Z}_{s}=Z_{s}+1/iC_{s}
\omega $, and $C_{x}=C(C+C_{c})/(2C+C_{c})$.

In terms of noise spectrum, the inelastic current through the asymmetric QDQ
is: %$$
%I=\frac{2e\pi T_c^2}{\hbar R_K }\frac{|Z(\Delta E
%/\hbar )|^{2}}{\Delta E^{2}}\left\{ S_{I}(\frac{\Delta E }{%
%\hbar })f_{L}(E_L)\left[ 1-f_{R}(E_R)\right] \right.
%$$
%\begin{equation}
%\left. -S_{I}(\frac{-\Delta E }{\hbar })f_{R}(E_R)\left[
%1-f_{L}(E_L)\right] \right\}.
%\end{equation}
\begin{equation}
I=\frac{2e\pi T_c^2}{\hbar R_K }\frac{|Z(\Delta E /\hbar )|^{2}}{\Delta E^{2}%
} (g_{LR}-g_{RL})
\end{equation}
where $g_{\alpha \beta}=S_{I}((E_\alpha-E_\beta)/\hbar) f_\alpha(E_\alpha)
(1-f_\beta(E_\beta))$. Notice that the output current of the subminiature
mesoscopic cell (i.e. the current through the appliance load $Z_{a}$) is
equal to the current through the DQD. In the following, we investigate in
detail the output current for the two schemes: (i) the noise originates from
the equilibrium noise offering circuit itself. (ii) the noise from the EMW
disturbance in a nearby external electromagnetic environment.

({\sl i}) {\sl the noise from the noise offering circuit}. In this case we
assume that the noise offering circuit is not affected by EMW disturbance in
the external environment; and the mesoscopic device is a quantum point
contact (QPC) in equilibrium. Then the noise spectrum $S_{I}(\omega )$ can
be expressed as:\cite{ref12,ref13} $S_{I}(\omega
)=4G\hbar\omega/(1-e^{-\hbar\omega/k_B T_{n}})$, where $G$ is the
conductance, $T_{n}$ is the temperature of the noise offering circuit (or
the QPC's system). % and $T_{n}$ is assumed not equal to $T_{D}$.
In the numerical studies, we take relevant parameters as: (1) $C=0.05fF$, $%
C_{c}=0.1fF$, $C_{s}=1nF$, and $Z_{s}=0.5R_{K}$; which are typical
experimental values.\cite{ref3} (2) Let $T_{c}=\Delta E /100$, so as to
satisfy the condition $\Delta E \gg T_{c}$. (3) Set $E_L=-E_R$ and $%
\mu_{L}=-\mu_{R}$. (4) Let the QPC only has one open channel, leading to $%
G=2/R_{K}$. First, we investigate the case in which the circuit is shorted,
i.e. $Z_{a}=0$, meanwhile $\mu_{L}=\mu_{R}=0$. The output current $I$ vs.
the temperature difference $\Delta T$ ($\Delta T\equiv T_{n}-T_{D}$) at
different $T_{D}$ is presented in Fig.2(a), showing an almost linear
relation. 
% At fixed $\Delta T$, the output current $I$ is reduced with increasing of
% $T_{D}$.
Fig.2(b) shows the output current $I$ vs the energy difference $\Delta E $,
exhibiting a non-monotonous variation. In both limits of $\Delta E 
\rightarrow 0$ and $\Delta E \rightarrow \infty $, $I$ becomes to zero. The
position $\Delta E $, at which $I$ has its maximum, is determined by $\Delta
T$ and $T_{D}$. It should be emphasized that $I$ is exactly zero if $\Delta
T=0$, regardless of the values of $T_{D}$ and $\Delta E $. This result is
obviously consistent with the second law of thermodynamics. In fact, one can
consider the cell as a heat engine, which works between two heat reservoirs
with the temperature $T_{n}$ and $T_{D}$.

({\sl ii}) {\sl the noise originates from the EMW disturbance in the nearby
environment}. In this case, we can assume that the temperature difference $%
\Delta T$ is zero, and the noise of the EMW disturbance is the only source
of the output current $I.$ For simplicity we also assume that the noise
spectrum $S_{I}(\omega )$ from the EMW disturbance is independent with $%
\omega $, i.e. a white noise, $S_{I}(\omega )=S_{0}$. The inset in Fig.2(b)
shows the output current $I$ vs. $\Delta E $ at the short circuit case. One
can clearly see that the output current $I$ is nonzero with the following
features: $I$ is zero at $\Delta E =0$, 
% due to the fact that the difference of electron's
% occupation number between the two levels $E_L$ and $E_R$
% is zero at $\Delta E =0$.
With the increase of $\Delta E $, the difference of the occupation number
becomes larger, leading to a larger current $I.$ Then $I$ reaches its
maximum when $\Delta E $ at a certain value (several times of $T_{n})$. 
% which is determined by $T_{n}$ or $T_{D}$ ($T_{n}=T_{D}$ in this section).
With the further increase of $\Delta E $, $I$ decreases slowly, because that
the transition from $E_R$ to $E_L$ becomes to very difficult at large $%
\Delta E $. It is merit mentioning that: in order to obtain a greater output
current $I$, one can take different ways as: (1) To lower the temperature of
the system, $T_{n}$ and $T_{D}$. (2) To change the noise offering circuit
into a LC circuit with the condition of $1/\sqrt{LC}=\Delta E /\hbar$.

Up to now, we have studied the output current $I$ at the short circuit case
(i.e. $Z_{a}=0$), in which the subminiature mesoscopic cell can produce a
current $I$, but its output power is zero. In the following, we study the
case of $Z_{a}\not{= }0$, in which the cell has a nonzero output power.
Since in this case, $\mu_{L}$ and $\mu_{R}$ are nonzero, the output current $%
I$ has to be solved from Eq.(3) and the equation $IZ_{a}=(\mu_{L}-\mu_{R})/e$%
. Fig.3(a) shows $I$ vs $Z_{a}$ for case (i). One finds that: (1) When $%
Z_{a} $ varies from $0$ to a critical value $Z_{ac}$, $I$ keeps almost
independent with $Z_{a}$. This means that this cell is a good constant
current source.\cite{add1} The critical appliance impedance $Z_{ac}$ is also
dependent with $\Delta E $, and is approximately determined by $%
Z_{ac}I=\Delta E/e$: $Z_{ac}\approx 3k\Omega$ and $Z_{ac}\approx 12k\Omega$
for $\Delta E =0.2meV$ and $\Delta E =0.3meV $, respectively. (2) When $%
Z_{a} > Z_{ac}$, the output current $I$ drops quickly with the increase of $%
Z_{a}$, quicker than $1/Z_{a}$.

The efficiency of the subminiature mesoscopic cell, $\eta ,$ is defined as
the ratio of the output energy of the cell to the input energy from the
high-temperature heat reservoir.\cite{ref14} Neglecting external losses, one
has $\eta =IV\Delta t/(I\Delta E \Delta t/e)=eV/\Delta E $, where $%
V=(\mu_{L}-\mu_{R})/e$ is the bias voltage across the appliance impedance $%
Z_{a}$. $\eta $ vs $T_{n}$ at different $Z_{a}$ is shown in the inset of
Fig.2(a). For comparison, the limit efficiency of heat engine, $%
(T_{n}-T_{D})/T_{n}$, is also shown by the dotted curve. The efficiency $%
\eta $ monotonously increases with the temperature $T_{n}$ and the appliance
impedance $Z_{a}$. Notice that $\eta $ can never excess $(T_{n}-T_{D})/T_{n}$%
, the efficiency of the ideal heat engine by the second law of
thermodynamics. But the efficiency $\eta $ may still be quite high, e.g. at $%
Z_{a}=10k\Omega $ and $T_{n}=3K$, $\eta \approx 57\%$; which is close to the
limit efficiency of heat engine, $66.7\%$.

Finally, let us study an interesting result. It is well known that for the
ordinary electric cell, the output current $I$ has its largest value if the
circuit is shorted. However, for the proposal cell, the output current $I$
may be larger if the circuit is loaded ($Z_{a}\not{= }0$) than the circuit
is shorted ($Z_{a}=0)$. Fig.3(b) shows $I $ vs. $Z_{a}$ at very low
temperature ($T_{n}=0.1K$ and $T_{D}=0.6K$). It is obvious that $I$ is not
the largest at $Z_{a}=0$. For example, for $\Delta E =0.2meV$, the largest
current $I$ is at $Z_{a}\approx 43k\Omega $, is not at $Z_{a}=0k\Omega $.
Because that the current $I(\Delta E)$ is directly proportional to the
modulus square of the trans-impedance, $\left| Z(\Delta E /\hbar )\right|^{2}
$, which also depends on $Z_{a}$ (See Eqs.(2) and (3)). While $Z_{a}$
increases, $\left| Z(\Delta E /\hbar )\right|^{2}$ increases quickly in a
certain range of $Z_{a}$, may lead to a larger current $I$ .

In summary, by using an asymmetric double quantum dots, we propose a
subminiature mesoscopic cell, which can change the the disordered noise
energy in a nearby mesoscopic device to the useful ordered electrical
energy. Two different cases of the origin of the noise are studied. The cell
manifests a good constant current source and high efficiency. Moveover, an
interesting phenomenon, that the output current may have its largest value
for the loaded circuit is discussed, which is important from the point of
view of the application.

%We gratefully acknowledge the financial support by the research grant
This work is supported by the research grant from the Chinese National
Natural Science Foundation, and the State Key Laboratory for Mesoscopic
Physics in Peking University, and a CRCG grant from the University of Hong
Kong.

%\bigskip

\section*{Figure Captions}

\begin{itemize}
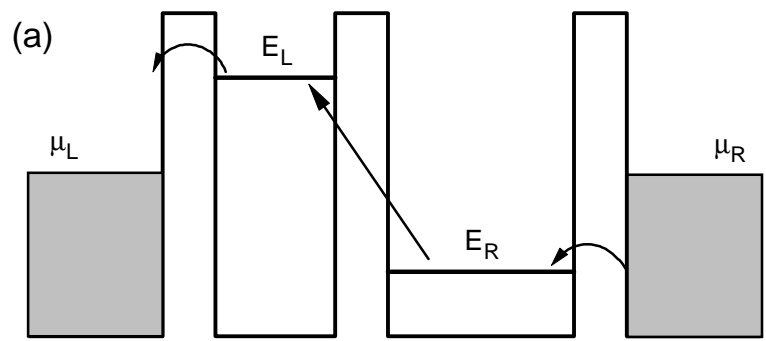
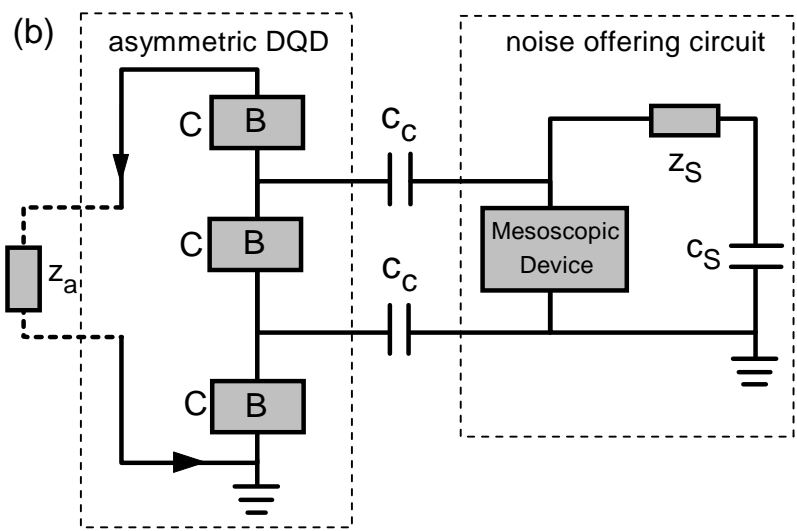
\item[{\bf Fig. 1}]  (a) Energy diagram of the asymmetric DQD. (b) A
schematic structure of the subminiature mesoscopic cell. The symbols marked
with "B" in the asymmetric DQD are the three tunnel barriers. $Z_{a}$ is the
appliance impedance. \vspace{4mm}

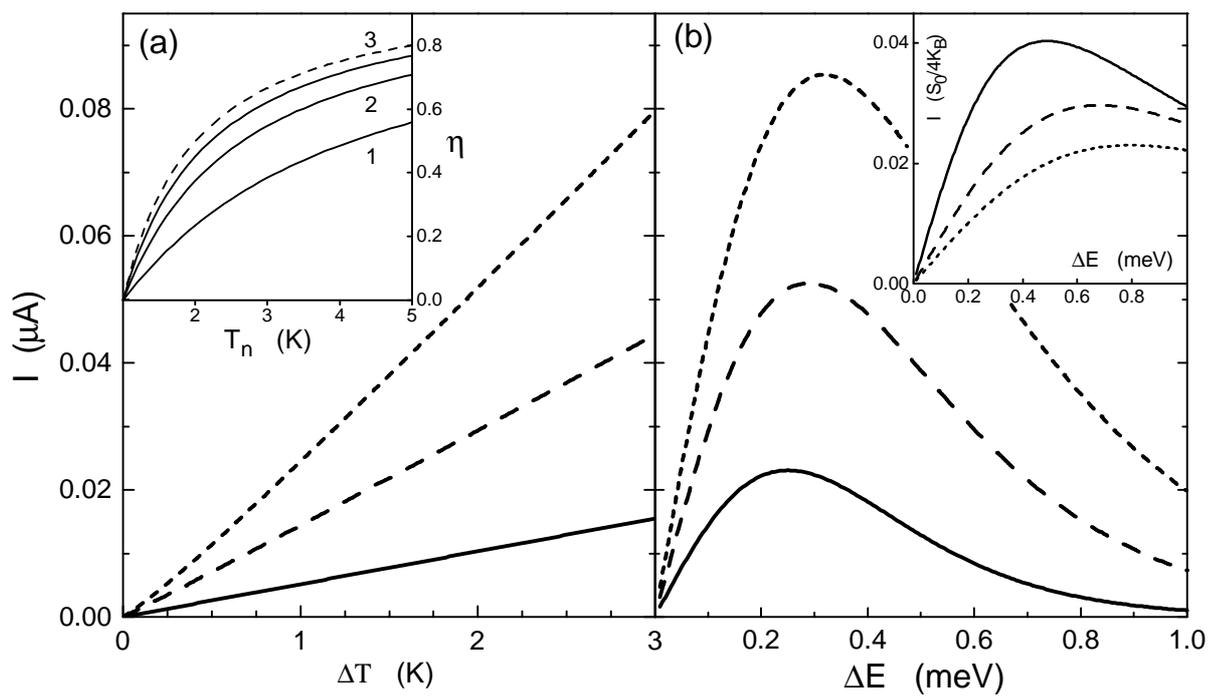
\item[{\bf Fig. 2}]  The output current $I$ of the short circuit for the
case of the noise from the oneself of the noise offering circuit. (a) $I$ vs 
$\Delta T$ at $\Delta E =0.1meV$. The dotted, dashed, and solid curves
correspond to $T_{D}=0.5K$, $1K$, and $3K$, respectively. (b). $I$ vs $%
\Delta E $ at $T_{D}=1K$. The dotted, dashed, and solid curves correspond to 
$\Delta T=3K$, $2K$, and $1K$, respectively. The inset in (a) shows the
efficiency $\eta $ vs $T_{n}$ at $T_{D}=1K$ and $\Delta E =0.1meV$. The
solid curves 1-3 correspond to $Z_{a}=3k\Omega $, $10k\Omega $, and $%
30k\Omega $, respectively. The dotted curve is $(T_{n}-T_{D})/T_{n}$ for
comparison. The inset in (b) shows the output current $I$ of the short
circuit for the case of the noise from the EMW disturbance in nearby
environment. The dotted, dashed, and solid curves correspond to $T_n=T_D=1K$%
, $2K$, and $3K$, respectively. \vspace{4mm}

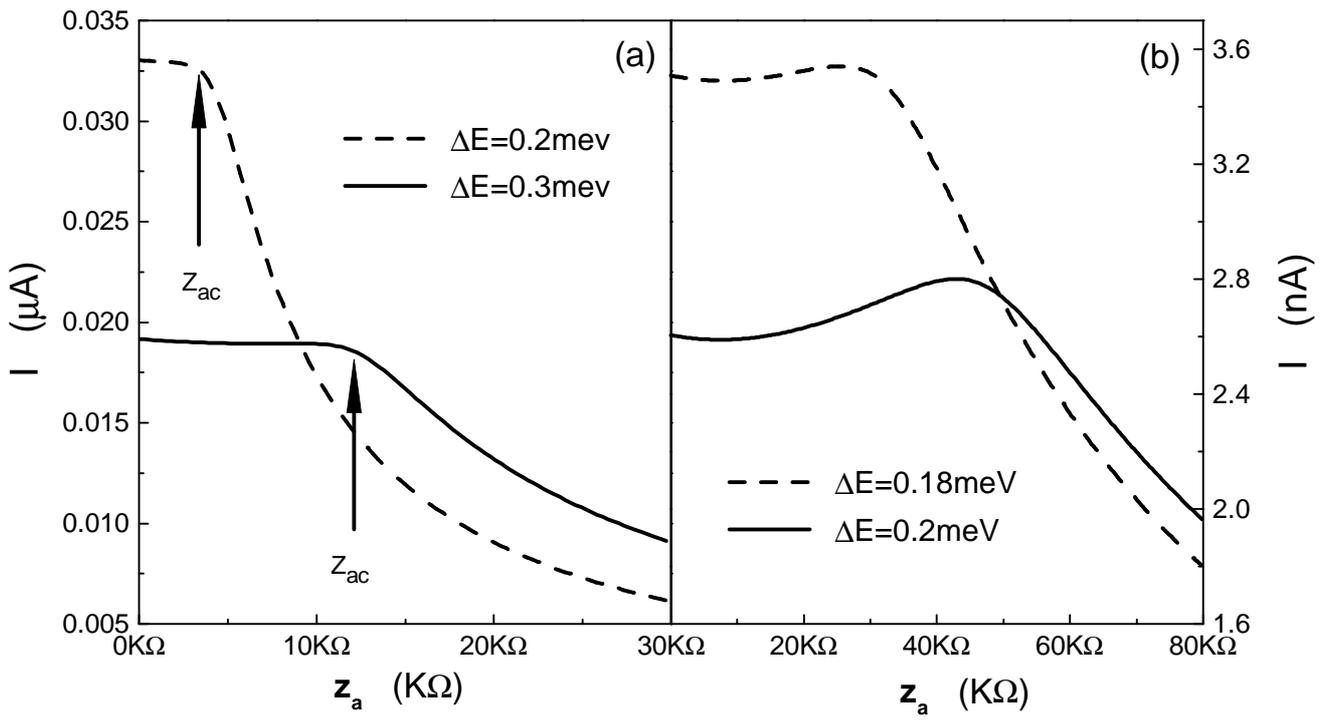
\item[{\bf Fig. 3}]  The output current $I$ vs $Z_a$. The parameters are:
(a) $T_{D}=0.1K$ and $T_n=1.5K$; (b) $T_{D}=0.1K$ and $T_n=0.6K$. \vspace{4mm%
}
\end{itemize}


\begin{references}
\bibitem{add1}  In order that the cell is a constant current source, one has
to maintain the constant temperatures $T_n$ and $T_{D}$ in experiment, which
need two heat reservoirs to touch the DQD system and the noise offering
circuit.
\end{references}
\end{document}